\pdfoutput=1 
\documentclass{JINST}
  \usepackage{multicol}
  \usepackage{float}

\title{Muonless Events in ICAL at INO}

\author{Ali Ajmi$^a$,
and S. Uma Sankar$^b$\\
\llap{$^a$}Homi Bhabha National Institute,\\
  Anushaktinagar, Mumbai 400 094, India\\
\llap{$^b$}Indian Institute of Technology Bombay,\\
  Mumbai 400 076, India\\
E-mail: \email{aliajmi@tifr.res.in},\\
  \email{uma@phy.iitb.ac.in}}

\abstract{The primary physics signal events in the ICAL at INO are the $\nu_\mu$ charged current (CC) interactions with a 
well defined muon track. Apart from these events, ICAL can also detect other types of neutrino interactions, i.e. the electron neutrino charged current interactions and the neutral current events. It is possible to have a dataset containing mostly $\nu_e$CC events, by imposing appropriate selection cuts on the events. The $\nu_\mu$CC and the neutral current events form the background to
these events. This study uses the Monte Carlo generated neutrino events, to design the necessary selection cuts to obtain
a $\nu_e$CC rich dataset. An optimized set of constraints are developed which balance the need for improving the purity
of the sample and having a large enough event sample. Depending on the constraints used, one can obtain a neutrino data sample, with the purity of $\nu_e$ events varying between 55\% to 70\%. 

 }

\begin{document}
 
\section{Introduction}
India-based Neutrino Observatory or the INO, is an experimental facility to be set up in the southern part of India. An important component of INO will be the Iron Calorimeter (ICAL). The ICAL aims to study the interactions of atmospheric neutrinos and antineutrinos. The determination of the neutrino mass hierarchy is one of its prime objects, apart from adding to the precision of the oscillation parameters \cite{Ghosh:2012px}.

ICAL is a giant magnetized neutrino detector, with Resistive Plate Chambers (RPCs) as the active detector elements \cite{Datar:2009zz,Athar:2006yb,Mondal:2012xm}. Efficient tracking abilities, good resolution of energy and timing, good identification of the charge of the particles are the essential capabilities of this detector   
 \cite{Riegler:2002vg,Lippmann:2003uaa}.

ICAL comprises of 3 modules, with $\sim$30,000 RPCs, and 151 iron layers weighing about 50kton in total. Each module contains 8 $\times$ 8 RPCs in a layer, each of which spans a surface of 1.84 m $\times$ 1.84 m and is 2.5 cm in width (height) \cite{Chatterjee:2014vta}. The steel structures to support them, however occupy gaps of 16 cm in between the RPCs. An additional pair of slits is also present in each module, making way for the current carrying coils of the magnet. The RPC layers are interspaced with iron plates of 5.6 cm thickness, and an air gap of 4 cm, thus making the successive RPCs 9.6 cm apart from each other in the vertical direction. 

The iron layers not only serve as the heavy target for most of the neutrinos but also carry the solenoidal magnetic field in the detector \cite{Behera:2014zca}.
 The neutrinos produce charged/neutral particles during interactions, which propagate through the detector. The iron layers are magnetised to $\sim$1.3 T. This accounts for the curvature in the tracks of the charged particles. The reconstruction of these tracks tells us the momentum and the charge of the particle. The kilometre-thick rock covering the INO cavern cuts down the cosmic ray particles. Hence, most of the events in the detector are produced by neutrino interactions only. Using the information of neutrino interactions in the detector, we can study the properties of these atmospheric neutrinos \cite{Thakore:2013xqa}.


The primary cosmic ray particles (like protons) interact with the atmospheric nuclei in the upper layers. To a large extent, these interactions produce pions, which eventually decay into muons and muon neutrinos. The muons can further decay, leading to both 
muon neutrinos and electron neutrinos in the final state. Effects of oscillations can be seen on the upward going neutrinos which  pass through the earth \cite{UmaSankar:2006yv}. Oscillations lead to the creation of the tau neutrinos in the neutrino flux.

 The neutrinos and anti-neutrinos undergo two types of interactions, depending on the mediating particle. The charged current (CC) interactions are mediated by the $W^{\pm}$ particles and the neutral current (NC) interactions by the $Z^0$. The CC events are different for each flavour and can be distinguished by the charged lepton in the final state. If the charge of this 
 lepton can be determined, then a distinction between neutrino and anti-neutrino interactions can be made. The NC events  
 have a  neutrino/antineutrino in the final state hence it is not possible to distinguish neither the flavour nor the neutrino from anti-neutrino in these events \cite{Balaji:1998rw}.


The CC interactions can be quasi elastic (QE), resonance scattering (RS) or deep-inelastic scattering (DIS) type. The QE interaction gives a lepton corresponding to the neutrino flavor. The RS interaction produces one/two pions too in addition to the lepton. The DIS interaction, which dominates beyond the neutrino energy of 4 GeV, gives a shower of hadrons apart from the charged lepton. 

The NC interactions can be of RS or DIS type only.
The RS and DIS interactions produce hadrons and a neutrino which goes undetected. So, only the hadron showers are visible to the detector in these cases.
In the case of NC elastic
scattering, because of no hadron shower in the final state, the event cannot be detected.

The presence of the thick iron layers puts a lower threshold on the minimum energy of a detectable particle. Therefore, the detection probability of sub-GeV neutrinos is very less, due to the trigger criteria, which removes most of the random noise events. 
 The neutrinos in the intermediate energy range of all the three flavors may interact with the detector.

  The ICAL is more efficient in the study of muon neutrinos in the GeV range. The presence of the magnetic field further enables us to tell apart the $\mu^-$ tracks from the $\mu^+$ tracks in the detector. However, apart from these well-recognizable muon track events, i.e., $\nu_\mu/\bar\nu_\mu$CC interactions, the ICAL will also contain  $\nu_e/\bar\nu_e$CC interactions and also the NC interactions of all three flavors \cite{BeckerSzendy:1992hq}. These interactions do not give clear muon tracks like the $\nu_\mu$CC events. Therefore, apart from the muon track containing events, we should also focus on these ``muonless'' events, in order to extract maximum possible information derivable from the ICAL detector.
  
  The paper is organized as follows. In section 2,  the details about the generated data which we have used in the study are explained, along with the details of how an event is detected and characterised. Section 3 deals with the application of various selection criteria which aim to increase $\nu_e$CC purity in the sample. The criteria are explained in detail, along with their physics justification. Their effects on the total dataset are included in the corresponding subsections. Some of the criteria are logically applied on the total dataset to check if the percentage of NC events can be enhanced, in section 4. Section 5 discusses the background contribution of the $\nu_e$CC and the NC events to the $\nu_\mu$CC events sample. In section 6, we have studied the contribution of the muonless events in determining the neutrino mass hierarchy. Section 7 concludes with a summary of the results and a few important comments.

Very few experiments have been able to study $\nu_e/\bar\nu_e$ events at higher energies. IMB, Kamioka, Super-Kamioka and very recently MINOS and T2K data are perhaps the only examples. ICAL is  sensitive to neutrinos in the energy range of  GeVs. Therefore, obtaining a source of data rich in electron neutrinos in this higher energy range is an important  aspect for the whole neutrino community. One will be able to study the characteristics of these high energy $\nu_e/\bar\nu_e$s from the atmosphere. 
Other possible physics studies are contribution of these events in the determination of the neutrino Mass Hierarchy, and also   utilization of the NC events in searches for the sterile neutrinos \cite{Giunti:2007ry,Kopp:2013vaa}.

\section{Simulations}
The following analysis is done with neutrino events generated by Nuance neutrino generator.
To reduce the influence of statistical fluctuations, we have generated 500 years of data. To begin with, we use normal hierarchy parameters. Simulations with inverted hierarchy parameters have also been checked, but not mentioned here, to avoid repetition.  The generated events are then simulated in the ICAL detector using GEANT4 \cite{Redij}.
 
\subsection{Signal-Detection}
Charged particles produced by the neutrino interactions pass through one or more RPCs and generate {\it hits}. These
hits are our primary signals. The layer number of RPC gives the z-coordinate of the hit. The x and y-coordinates are
given by the copper-strips of the pick-up panels which are orthogonally oriented at the top and 
the bottom of the RPCs. The number of strips in x-direction, with a signal, gives x-strips and similarly in y-direction gives y-strips.
The maximum of the number of x-strips or y-strips is defined to be the number of ``strips-hit'' in that layer. This number of strips-hit in a layer, when summed over all the layers which have received hits in that event, gives the number of strips-hit in that event. The hits distribution mentioned  hereafter, refers to this value of the number of strips-hit in an event.

\subsection{Types of events and their signatures}
Vertical and high energy muons travel through large number of layers before stopping/decaying. Therefore, the $\nu_\mu$s which have high energy and are incident mostly along the vertical direction give hits in more number of layers, in case of CC interactions. In fact, the muons thus produced, form clear tracks in the detector and their momentum can be reconstructed. The curvature induced along the muon path due to the magnetic field, leads to the charge identification. On the contrary, the $\nu_\mu$s which have lower energy or are incident mostly along the horizontal direction \cite{Fukuda:1998ub} are confined in less number of layers. They can hardly be distinguished from the hadron showers which also emanate at the event vertex.

The $\nu_e$CC events produce electrons which can give rise to em showers, but no track can be seen.
The NC events have no charged lepton in the final state and hence have lower number of hits.  They have only a set of
hadrons in the final state and, in general, are indistinguishable from $\nu_e$CC events. \cite{Hasert:1973cr,Hasert:1973ff}

 \section { Selection Criteria to obtain $\nu_e$CC rich Sample:}
 We devise certain conditions to ensure that the selected event sample contains mostly $\nu_e$CC events, with minimum possible background of $\nu_\mu$CC and NC events.
 
 The simplest information available to us about an event is: (i) the number of hits that an event gives in the detector and, (ii) the number of layers it is passing through. Therefore, one first focuses on them.
 
 In order to understand the behaviour of the different neutrino events in the detector, we first have a look at the nature of hits distributions of all the three event types: the NC, the $\nu_e$CC and the $\nu_\mu$CC types, in different energy ranges of the incident neutrinos.
 \newpage
 \vskip -8ex
   	\begin{figure}[H]	
 \centering
 \setlength\fboxsep{0pt}
 {\includegraphics[width=1.\textwidth,height=0.249\textheight ]{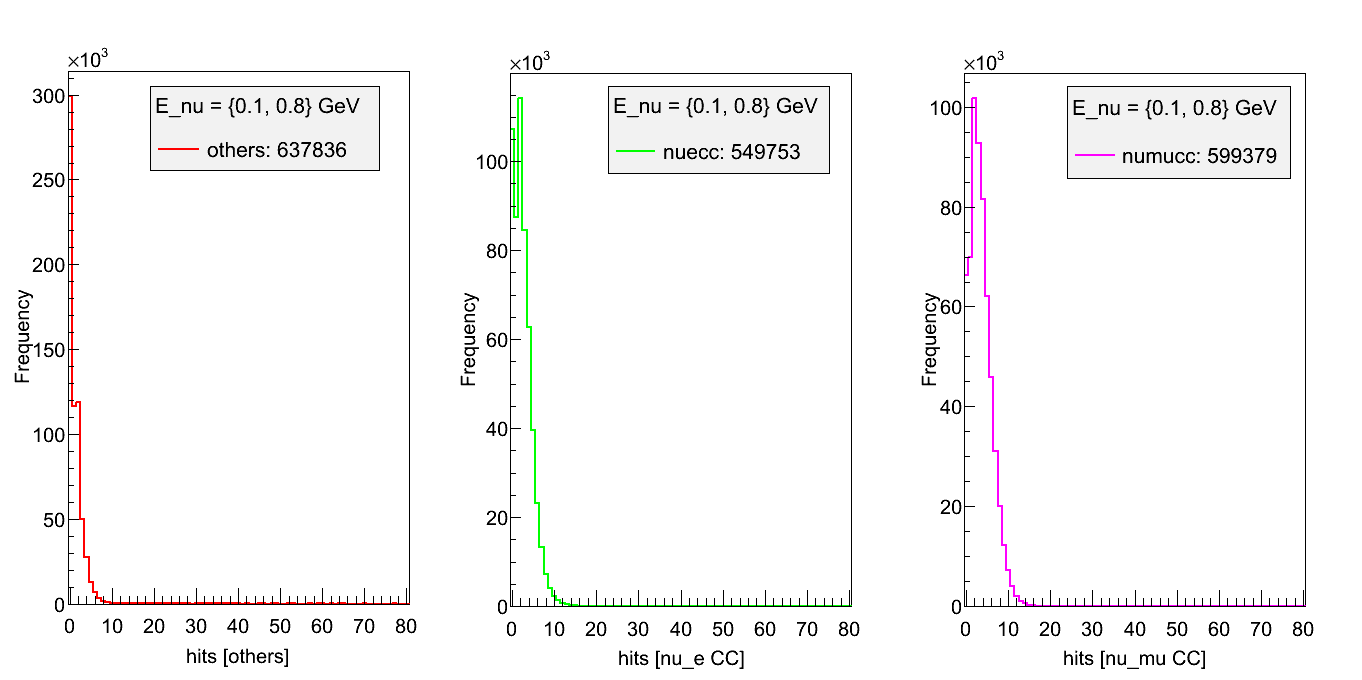} }
\end{figure}
\vskip -12ex
   	\begin{figure}[H]	
 \centering
 \setlength\fboxsep{0pt}
 {\includegraphics[width=1.\textwidth, height=0.249\textheight]{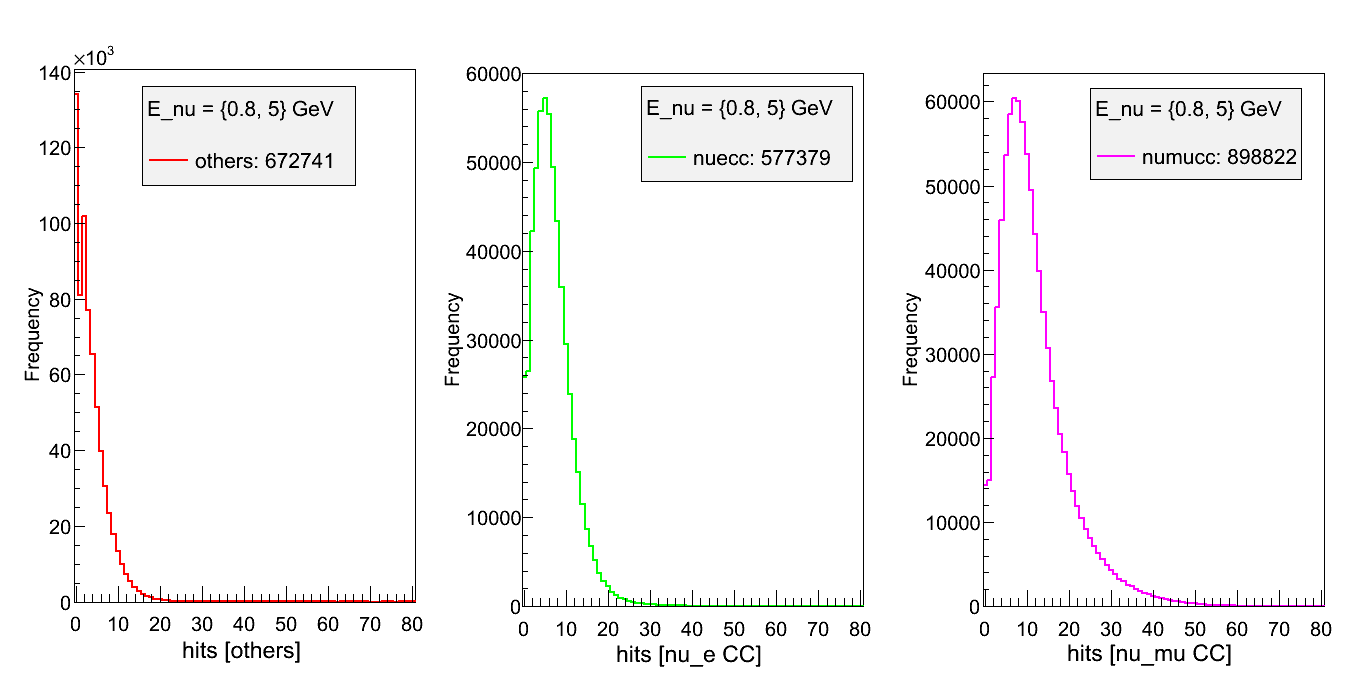} }
\end{figure}
\vskip -12ex
   	\begin{figure}[H]	
 \centering
 \setlength\fboxsep{0pt}
{\includegraphics[width=1.\textwidth, height=0.249\textheight]{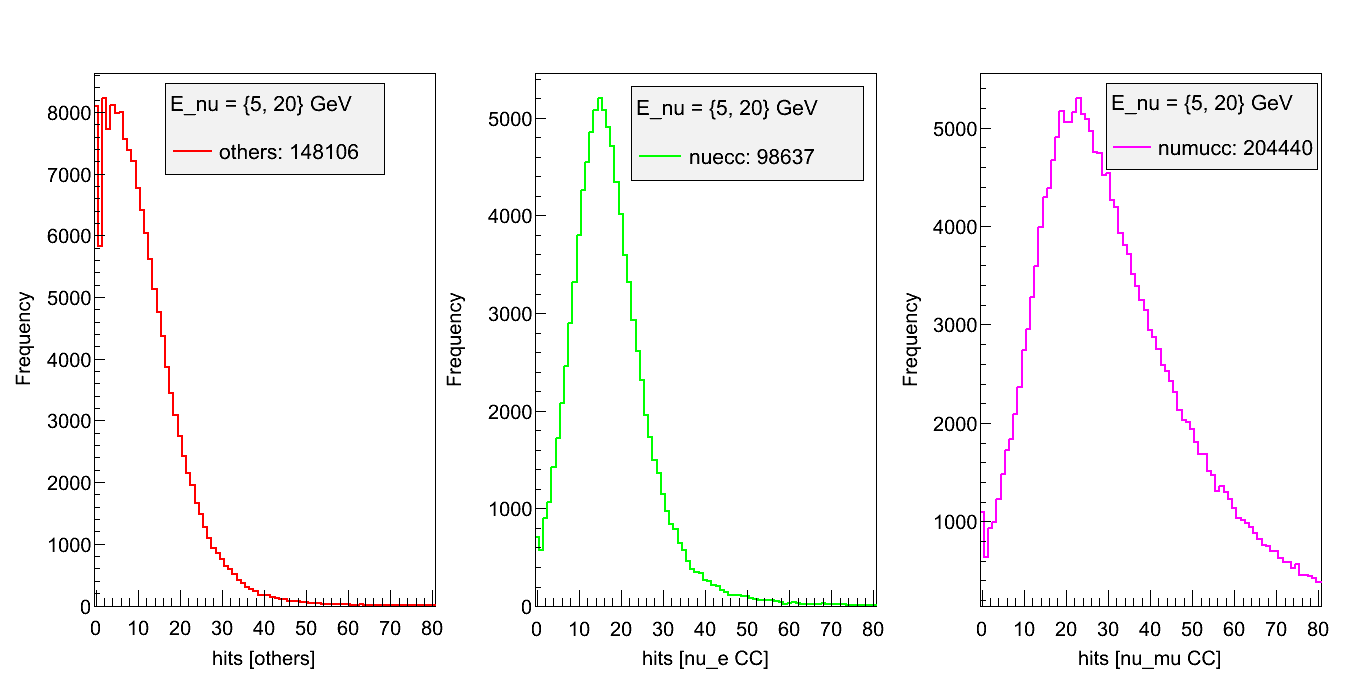} }
\end{figure}
\vskip -12ex
   	\begin{figure}[H]	
 \centering
 \setlength\fboxsep{0pt}
{\includegraphics[width=1.\textwidth, height=0.249\textheight]{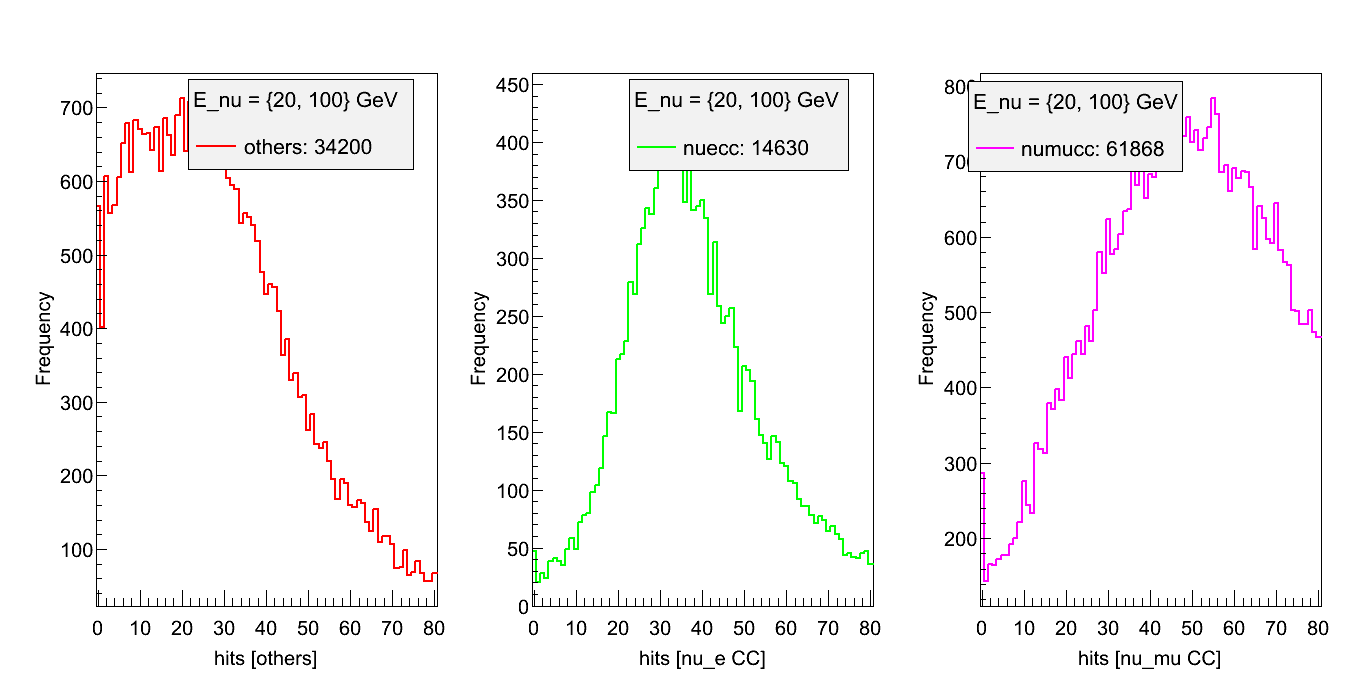} }
\vskip -5.5ex
 \caption{\small Hits distribution in the neutrino energy bins (from top to bottom in order): E$_\nu$=\{0.1,0.8\} GeV; E$_\nu$=\{0.8,5.0\} GeV; E$_\nu$=\{5.0,20.0\} GeV; E$_\nu$=\{20.0,100.0\} GeV, in case of the three types of neutrino events (from left to right in each row): $\nu_e$CC; all NC (+$\nu_\tau$CC); $\nu_\mu$CC event types. The x-axis gives the number of hits, and y-axis contains the counts of the events. }

\end{figure}
 
 The distributions in figure 1 show that, for $\nu_\mu$CC events, the number of hits is greatly enhanced with increasing energy. This increase is much less for $\nu_e$CC events and hardly exists in case of the NC events. The figure also clearly suggests a lower threshold of $\sim$10 hits to suppress a large fraction of NC events and low energy $\nu_e,\nu_\mu$CC events. With this cut, 12\% of the total NC events is retained. The survival fraction for $\nu_e$CC events is about 18\%. So, out of the total set of survived data of the $\nu_e$CC and the NC, more than 60\% are $\nu_e$CC events, as seen in
 Table~\ref{Table1}. The events containing high energy muons can be separated by restricting the number of layers or by being identified by the track reconstruction algorithm of ICAL.

Below, we describe the selection criteria to generate event samples with maximum $\nu_e$CC events.


\subsection{Hits and Layers}

 The nature of the hits distributions is observed for all the three types of interactions mentioned above.
A selection cut on the number of hits appears to be  an effective way to obtain an events sample with a majority of $\nu_e$CC events. The NC events peak at very low number of hits, unlike the CC events. So, it is difficult to select an event sample rich in NC events. 

The electrons/positrons, in general, travel a shorter distance than the hadrons. On the other extreme, the muons of the $\nu_\mu$CC events travel through several layers. A primary observation of the layer distribution shows that a cut on the number of layers hit in an event is an effective criterion. Here the ``layers'' refer to the number of those layers which receive one or more hits in an event. However, the layer cut is a very sensitive cut, owing to the thick iron layer in between two RPC layers.

As mentioned earlier, $\nu_\mu$CC events either with low energy muons and/or in horizontal direction, do not have identifiable muon track. Such events have been found to be a significant contributor in the selected events sample. So, a separate count is maintained for them, in order to make it easy to derive a strategy to reduce them. The rest of the backgrounds, i.e., the $\nu_e$NC, $\nu_\mu$NC, $\nu_\tau$CC and NC are all contained in the $``$others$"$. The $\nu_\tau$CC events being pretty small in number, are not separately counted.
 	\begin{figure}[H]	
 \centering
 \setlength\fboxsep{0pt}
 {\includegraphics[width=.9\textwidth]{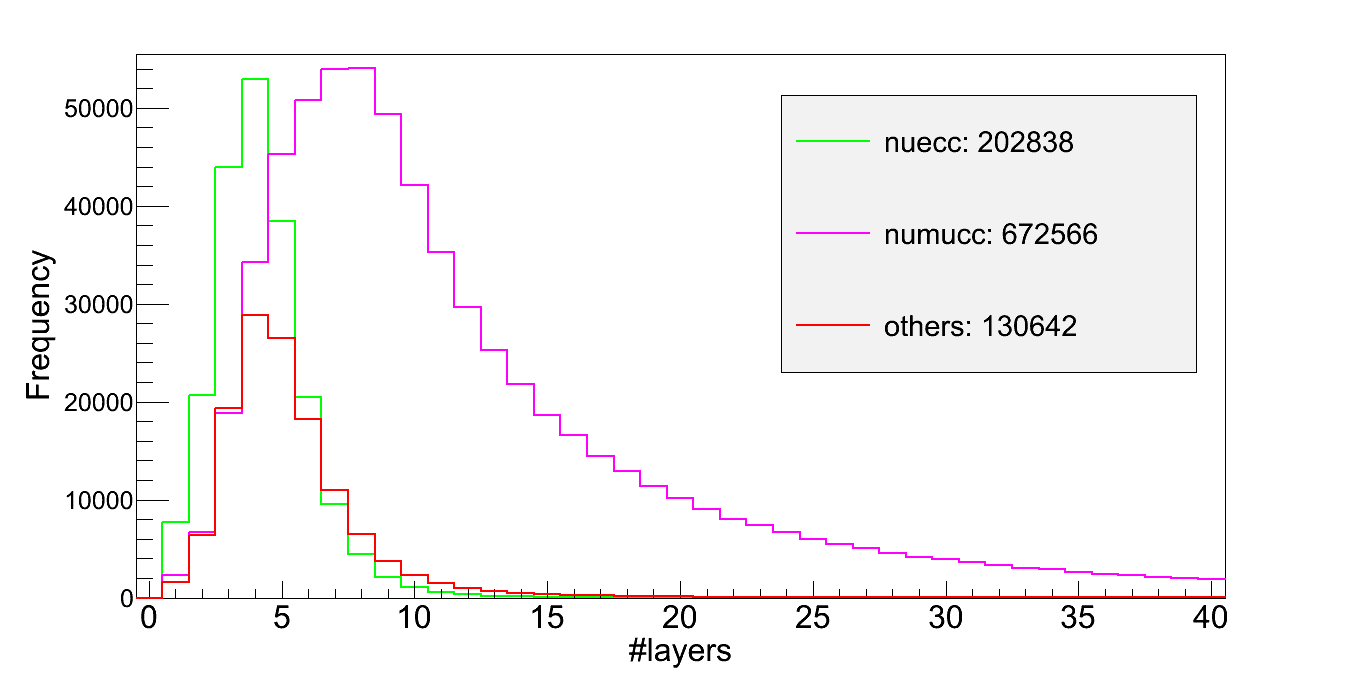} }
 \caption{Distribution of number of layers for all events with more than 10 hits, for the 500years NH data in E$_\nu$= \{0.1,100\} GeV.   }
\end{figure}

 The $\nu_e$CC events layer distribution, after a minimal hits cut (say 10), peaks around 5 while that of the $\nu_\mu$CC events peaks around 10 layers. So, by selecting events which are confined in 5 layers or less, we can reject 
$\nu_\mu$CC events  which are energetic and/or vertical. However, low energy $\nu_\mu$CC events, especially those in
the horizontal direction, do pass this cut and give rise to the events listed below.
 Various cuts on the number of hits and layers have been imposed on the set of events ($E_\nu$=\{0.1,100\} GeV). 
 A few significant ones among them are tabulated in Table~\ref{Table1}.
 

  	\begin{table}[H]		
 \centering
\begin{tabular}{|p{3.8cm}|p{2.5cm}|p{2.5cm}|p{2.5cm}|}
 \hline
  \textbf{Selection Criteria}&	{\textbf{$\nu_e$CC }}&{\textbf{others }}&{\textbf{$\nu_\mu$CC }}\\
 \hline
  \textbf{hits$>$0 }	&	1106742	&	1050814	&	1682527	\\
 \hline
  \textbf{hits$>$10 (only)}	&	202838	&	130642	&	672566	\\
 \hline
  \textbf{hits$>$15 (only)}	&	97535	&	69340	&	445977	\\
 \hline
  \textbf{hits$>$20 (only)}	&	52398	&	42476	&	314597	\\
 \hline
  \textbf{hits$>$15; 
	layers$\leq$4}	&		47711&		19390&		19875\\
 \hline
  \textbf{hits$>$15; 
	layers$\leq$5}	&		68702&		32953&		36211\\
 \hline
  \textbf{hits$>$15; 
	layers$\leq$7}	&		89614&		52550&		76194\\	
 \hline
  \textbf{hits$>$10; 
	layers$\leq$4}	&		125321&		56177&		62113\\
 \hline
  \textbf{hits$>$10; 
	layers$\leq$5}	&		163807&		82717&		107350\\
 \hline
 \end{tabular}
\caption{Events counts after applying the selection cuts on the Geant output of the NH 500 years data in E$_\nu$=\{0.1,100\} GeV.}
\label{Table1}
\end{table}

%

Very high number of hits are mostly due to $\nu_\mu$CC events. So, an upper cut on hits can eliminate such events.

 \paragraph{Average hits per layer criteria:}
This is a criterion derived by combining the above two, but still may help in eliminating the track containing events. 
The muon tracks give mostly 2-3 hits in a layer. So, applying a lower cut on the average hits per layer (hpl) seems to be quite reasonable in rejecting most $\nu_\mu$CC events. 
The hits per layer cut is useful in studying vertical, high energy muon events. Here, 
given that the number of layers is constrained to be $\leq 5$, it is not very effective.
The addition of this cut leads only to a marginal improvement. 

Applying the above and more similar cuts, it is seen that the fraction of $\nu_e$CC events in the sample increases, but at the cost of sample size and fraction of vertical events. Hence, an optimized set of criteria needs to be chosen from the varied
sets of cuts mentioned here.

 \subsection{Distribution pattern of the hits in the layers:}
The cuts on the basis of hits and layers are indeed the simplest and very effective selection criteria. However, a number of various other parameters have also been studied, to ensure how much they can contribute to improving the purity of the $\nu_e$CC events in the sample. (Please refer to INO internal note \cite{internal} for details.)

%

The behaviour of an $\nu_e$-interaction is certainly different from the other two types of interactions, as far our physics knowledge is concerned. The presence of the electron/positron makes it stand apart from the NC interactions. The way electrons/positrons lose their energies in the detector is different from the way $\mu^+/\mu^-$ do. The challenge 
is to utilize these characteristics in distinguishing $\nu_e$ events from $\nu_\mu$ events in the data from the ICAL detector. 

We have so far observed the overall picture of the number of hits and layers in an event. As shown earlier in this paper, these cuts have left us with enough hope to invest more interest in this study.  So, it is required that we look deeper, to decipher those characterizable features of the $\nu_e$CC, which can be readily realised from the ICAL data.

\subsubsection{Maximum Hits Difference }
The $\nu_e$CC events contain em showers. These should have generated a huge number of hits, but most of them are absorbed by the thick iron layers. However, in  some events the shower may start at the verge of the iron layer. In such cases, a sudden and significant increment in the number of hits in the following layer is expected. 

The difference in the number of hits in two adjacent layers in an event is calculated. This difference is maximized over all such pairs in that event. The value of the maximum difference in hits thus obtained forms our present selection criterion. 
The effect of this cut is shown in Table~\ref{Table2}.

\begin{table}[H]		
 \centering
\begin{tabular}{|p{4.8cm}|c|c|c|}
 \hline
  \textbf{Selection Criteria}	&{\textbf{$\nu_e$CC }}&{\textbf{others}}&{\textbf{$\nu_\mu$CC }}\\
 \hline
  \textbf{h$>$10; 
	L$\leq$5; }		&		163807&		82717&		107350\\
 \hline
  \textbf{h$>$10; 
	L$\leq$5;
	max hits diff.$>$5}	&		82500&		34701&		38824\\
 \hline
  \textbf{h$>$15; 
	L$\leq$5; }		&		68702&		32953&		36211\\
 \hline
  \textbf{h$>$15; 
	L$\leq$5;
	max hits diff.$>$5}	&		50295&		21844&		23991\\
 \hline
 \end{tabular}
\caption{Events counts after applying the hits-layers selection criteria and adding the cut on maximum difference in the number of hits in adjacent layers. (500 years NH data in E$_\nu$=\{0.1,100\}GeV.)[{\bf ``h''=\#hits; ``L''=\#Layers }] }
\label{Table2}
\end{table}

This selection criterion adds to improving the $\nu_e$CC events ratio by about 3-4\%. However, a simultaneous study of the known (Nuance) information shows that the larger hits difference is given by mostly horizontal $\nu_e$CC events. 

\subsubsection{Comparing the hits in each layer:}
 
 The number of hits in every individual layer in an event is studied.  This criterion, in a way seeks a pattern in the number of hits in adjacent layers. A variety of patterns are assumed and checked with the set of events. Two of them are stated below. The underlying logic still rests on the concept of the em shower.

 \textbf{Additional hits in the next layer:} One of the layers hit is chosen and additional 5 or 6 hits are demanded in the very next layer. All the layers in that event are also checked. The event to be selected must have at least one such a pair of layers. It is to be simultaneously noted that a lower threshold of 2 layers becomes inherent.

\textbf{Majority of hits in one layer:} One can call this criterion a modified version of the earlier one. According to this criterion, the event must contain 50\% or 60\% of the total number of hits in a single layer. Therefore, no lower cut on the number of layers is required here. 

The effect of the selection cuts are tabulated in Table~\ref{Table3}.
\begin{table}[H]		

 \centering
\begin{tabular}{|p{5.2cm}|c|c|c|c|}
 \hline
  \textbf{Selection Criteria}	&{\textbf{$\nu_e$CC }}&{\textbf{others}}&{\textbf{$\nu_\mu$CC }}&{$\nu_e$CC purity \%}\\
 \hline
 \hline
  \textbf{hits$>$15; 
	layers$\leq$5; }	&		68702&		32953&		36211&		50\\
 \hline
  \textbf{hits$>$15; 
	layers$\leq$5; 
	h$_L>$h$_{L\pm1}$+5}	&		47009&		21191&		22934&		52\\
  \hline
  \textbf{hits$>$15; 
	 layers$\leq$5; 
	h$_L>$50\% hits}	&		38479&		13745&		16934&		56\\
  \hline
  \textbf{hits$>$15; 
	 layers$\leq$5; 
	h$_L>$60\% hits}	&		29123&		9038&		11948&		58\\
  \hline
  \hline
  \textbf{hits$>$15; 
	 layers$\leq$4; }	&		47711&		19390&		19875&		55\\
 \hline
  \textbf{hits$>$15; 
	 layers$\leq$4; 
	h$_L>$h$_{L\pm1}$+5}	&		34399&		13308&		13868&		56\\
  \hline
  \textbf{hits$>$15; 
	 layers$\leq$4; 
	h$_L>$50\% hits}	&		32737&		10931&		12679&		58\\
  \hline
  \textbf{hits$>$15; 
	 layers$\leq$4; 
	h$_L>$60\% hits}	&		26006&		7735&		9690&		60\\
\hline
 \end{tabular}
\caption{Events counts after applying the hits-layers selection criterion and demanding (i) 5 additional hits in adjacent layers (h$_L$, h$_{L\pm1}$); (ii) 50-60\% of total number of hits in one layer. (500 years NH data in E$_\nu$=\{0.1,100\} GeV.) [{\bf ``hits''=total \#hits; ``h$_L$''=hits in any of the layers, say the L$^{th}$ layer.}] }
\label{Table3}
\end{table}

\subsubsection{The Overall Distribution Pattern of Hits from the average number of hits among the layers} 
The hits in different layers of the $\nu_e$CC events are non-uniform. The hits are mostly over concentrated in some layers, while entirely sparse in the rest. 
 	\begin{figure}[H]	
 \centering
 \setlength\fboxsep{0pt}
 \fbox{\includegraphics[width=.9\textwidth]{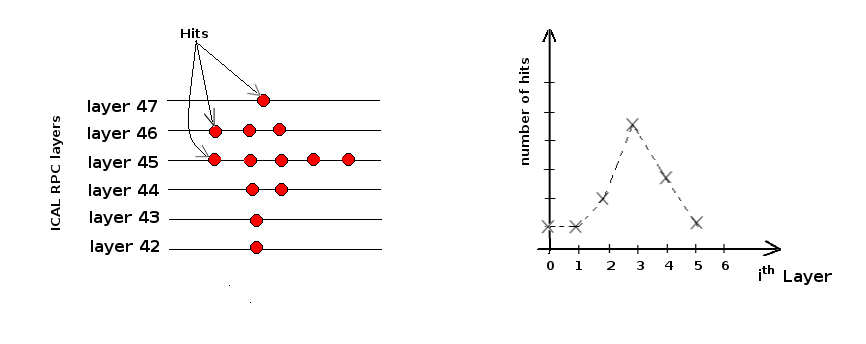}}
 \caption{Schematic Diagram of hits in the RPC layers}
\end{figure}
Figure 3 shows, the hit pattern among various layers in an event (left panel) and the number of hits
vs layer number (right panel). For the plot in the right panel, the lowest layer hit is labelled to
be {\bf 0}, the next layer is {\bf 1} and so on. In such a plot, the $\nu_\mu$CC gives a broader peak than the $\nu_e$CC / NC. Hence, events selected with such sharper peaks should reduce the fraction of $\nu_\mu$CC events in our sample. This property can be parametrized as either the mean or RMS value of the layerwise hits distribution of each event. However, having studied both the quantities, the cut on the RMS value appeared comparatively more effective. Some of the results are shown in Table~\ref{Table4}. 

\begin{table}[H]		
 \centering
\begin{tabular}{|p{6.5cm}|c|c|c|p{1.5cm}|}
 \hline
  \textbf{Selection Criteria}	&{\textbf{$\nu_e$CC }}&{\textbf{others}}&{\textbf{$\nu_\mu$CC }}&{$\nu_e$CC purity \%}\\
 \hline
  \textbf{h$>$15; 
	L$\leq$5; }	&			68702&		32953&		36211&		50\\
 \hline
  \textbf{h$>$15; 
	L$\leq$5;
	rms$<$1.2	}	&		56254&		24916&		25431&		53\\
 \hline
  \textbf{h$>$15; 
	L$\leq$5;
	rms$<$1.2	;
	max hits diff.$>$4}	&		48248&		20452&		21241&		54\\
 \hline
  \textbf{h$>$15; 
	L$\leq$5;
	rms$<$1.2	;
	max hits diff.$>$6}	&		39610&		15585&		16969&		55\\		
 \hline
  \textbf{h$>$10; 
	L$\leq$4; }	&			125321&		56177&		62113&		51\\		
 \hline
  \textbf{h$>$10; 
	L$\leq$4;
	  rms$<$1.2	}	&		111858&		47961&		52860&		53\\	
 \hline
  \textbf{h$>$10; 
	L$\leq$4;
	  rms$<$1.2;
	max hits diff.$>$3}	&		86157&		35115&		37026&		54\\		
 \hline
  \textbf{h$>$10; 
	L$\leq$5;
	  rms$<$1.2;
	max hits diff.$>$3}	&		99814&		43409&		46455&		56\\		
 \hline
  \textbf{h$>$10; 
	mean$<$2;
	  rms$<$1.2;
	max hits diff.$>$3}	&		83954&		35130&		36127&		54\\		
  \hline
  \textbf{h$>$10; 
	mean$<$2;
	  rms$<$1.2;
	max hits diff.$>$5}	&		60959&		23063&		24129&		56\\		
 \hline
  \textbf{h$>$10; 
	mean$<$2;
	  rms$<$1.2;
	max hits diff.$>$5;
	hpl$>$4}	&			51249&		18247&		18922&		58\\		
 \hline
 \end{tabular}
\caption{Events counts after applying the hits-layers selection criterion; adding the cut on the variance from the mean of the vertical distribution of hits in layers, i.e. rms; the criteria of max hits diff.  included for further improvement. (500 years NH data in E$_\nu$=\{0.1,100\} GeV.) [{\bf ``h''=\#hits; ``L''=\#Layers; ``hpl''=avg hits/layer.}]}
\label{Table4}
\end{table}

The RMS cuts appear quite effective in improving the ratio. In fact, the cut of maximum difference in the hits  further adds to bettering the results. Therefore, one can obtain $\sim$55\% majority of $\nu_e$CC events, with a moderately large sample size.

 \section{The NC events fraction}
 Our primary/main focus in this paper is to obtain a $\nu_e$CC rich sample of neutrino events at ICAL. So, the selection cuts so far have been favoring $\nu_e$CC events. However, the NC events fraction can also be enhanced comparatively. 
 
 The cuts are based on the simplest criteria of hits and layers. However, the extent of this possibility depends on the hardware threshold to be put in the ICAL, for accepting an NC event. This requires a trigger algorithm different from that for the muon track-containing events. 
 	\begin{figure}[H]	
 \centering
 \setlength\fboxsep{0pt}
{\includegraphics[width=0.8\textwidth ]{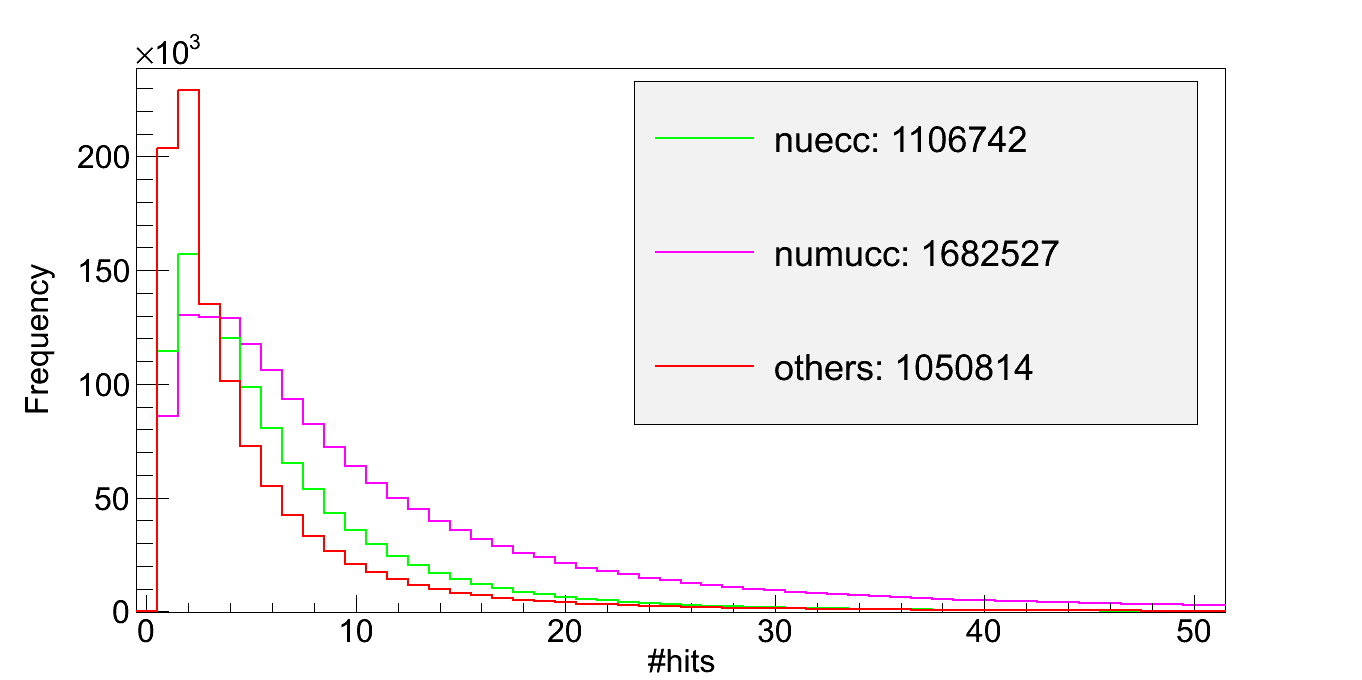}}
 \caption{Distribution of number of hits for all non-zero hit events with E$_\nu$ =  \{0.1,100\} GeV for the 500years NH data.   }
\end{figure}

	\begin{figure}[H]	
 \centering
 \setlength\fboxsep{0pt}
{\includegraphics[width=0.8\textwidth ]{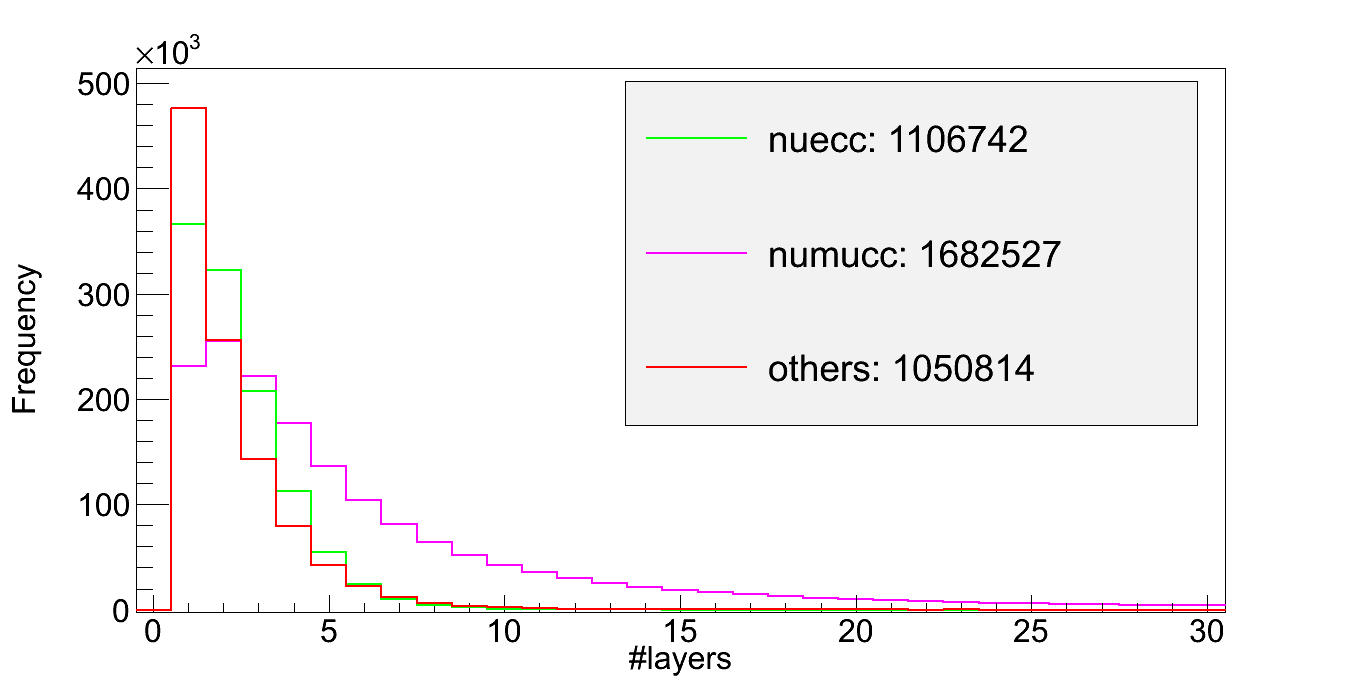}}

 \caption{Distribution of number of layers which received one or more hits in an event, E$_\nu$ =  \{0.1,100\} GeV for the 500years NH data.   }
\end{figure}

The NC events give mostly very less number of hits, and are confined in very few layers, as in figure 4 and 5. So, one might be tempted to put an upper threshold on the number of hits, to obtain an NC events rich sample. But, to deal with such a small number of hits in an event, one must be aware of neutrinos with sub-GeV energies too. So, the dataset including neutrinos with E$_\nu$ =  \{0.1,100\} GeV is satifactory to proceed further.  
  

 	\begin{table}[H]		
 \centering
\begin{tabular}{|p{3.5cm}|p{2.5cm}|p{2.5cm}|p{2.5cm}|p{1.5cm}|}
 \hline
  \textbf{Selection Criteria}&	{\textbf{$\nu_e$CC (NH evts)}}&{\textbf{others (NH evts)}}&{\textbf{$\nu_\mu$CC (NH evts)}}&{NC purity \%}\\
   \hline
   \hline
  \textbf{0$<$hits$\leq$10 (only)}&		903904	&	920172	&	1009961	&	32\\
 \hline
  \textbf{0$<$hits$\leq$10; 
	  layers$\leq$2}&			659926	&	724065	&	478480	&	39\\
   \hline
  \textbf{0$<$hits$<$4 (only)}&		406705	&	568177	&	345322&		43	\\
 \hline
  \textbf{0$<$hits$<$4; 
	layers$\leq$2}	&		397895&		558109&		321648&		44\\
 \hline
  \textbf{0$<$hits$<$4; 
	layers = 1}	&		287799&		436263&		198539&		47\\
 \hline
  \textbf{4$\leq$hits$\leq$10; 
	layers = 1}	&		70330&		38360&		31046&		27\\
\hline
\end{tabular}
\caption{Enhancing NC fraction: Events counts after applying the selection cuts on the Geant output of the NH 500years data files in E$_\nu$=\{0.1,100\} GeV}
\label{Table5}
\end{table}
The NC counts in the selected sample are almost equal to the sum of the selected $\nu_e$CC and the $\nu_\mu$CC events. 
The NC events have a very small number of hits in general. The cuts used earlier demanded a minimum of 10 hits and hence
discriminated against the NC events. To favour NC events, one should design a cut demanding a very small number of hits. The dominance of the NC events can be gradually realized in case of events having 10 hits or lower, as in Table~\ref{Table5}.

If we demand \# hits $\leq$ 3 and \# layers = 1 or 2, we get the event samples shown in the above table, which 
are quite rich in NC events. If the noise is kept under control, such events can be used to study mixing with
sterile neutrinos. Trigger efficiency will play a major role in selecting such events. In fact, it has been checked that a sample of single hit events has more than 50\% NC events in the sample. But obviously, just one-hit is an unacceptable criteria. Therefore, the selection cuts will have to be redesigned entirely, to obtain an events sample containing NC events in significant majority.

\section{$\nu_e$CC and NC events as background to $\nu_\mu$CC events}
 
Muon track containing events are the primary data for the ICAL, especially those within the range E$_\nu$=\{0.8,20\}GeV.
   The $\nu_\mu$CC events detected at ICAL must pass through (i.e. give hits in) a minimum number of layers (5 or 6), so that the muon track can be reconstructed. This layer cut will undoubtedly select mostly the $\nu_\mu$CC events, which form the signal events in this case. However, some $\nu_e$CC and NC events also will pass through this cut and form background to this events sample. 
  
  As shown in Table~\ref{Table6}, out of all the generated events, about 20\% of the ``others'' do not give any hits in the ICAL. For $\nu_e$CC and $\nu_\mu$CC events, this fraction of ``undetectable'' events is about 5\%. The layers distribution of these events (for the energy range E$_\nu$=\{0.1,100\}GeV) is shown in figure 5. 
   	\begin{table}[H]		
 \centering
\begin{tabular}{|p{3.5cm}|p{2.5cm}|p{2.5cm}|p{2.5cm}|}
 \hline
  \textbf{Selection Criteria}&	{\textbf{$\nu_e$ CC  }}&{\textbf{others (NC+$\nu_\tau$CC)}}&{\textbf{$\nu_\mu$ CC }}\\
 \hline
  \textbf{all generated events}	&	676014	&	820854	&	1103263	\\
 \hline
  \textbf{\#events with hits$>$0 in ICAL}	&	649487	&	678590	&	1087709	\\
   \hline
 \end{tabular}
\caption{Events counts before applying the selection cuts on the Geant output of the NH 500years data files in E$_\nu$=\{0.8,20\}GeV. }
\label{Table6}
\end{table}

 Since reconstructable $\nu_\mu$CC events demand a minimum number of layers to be hit, the distributions with two such layer-cuts are quantified in Table~\ref{Table7}. These basically fetch us the number of events or percentage composition of the selected events sample. This feature of large suppression of the $\nu_e$CC and NC events with this cut is evident in figure 5.

      	\begin{table}[H]		
 \centering
\begin{tabular}{|p{3.5cm}|c|c|c|}
 \hline
  \textbf{Selection Criteria}&	{\textbf{$\nu_e$ CC  }}&{\textbf{others (NC+$\nu_\tau$CC)}}&{\textbf{$\nu_\mu$ CC }}\\
 \hline
  \textbf{\#events: L$\geq$5}	&	84115	&	73849	&	683635	\\
 \hline
   	&	$\sim$10\%	&	$\sim$9\%	&	$\sim$81\%	\\
 \hline
 \hline
  \textbf{\#events: L$\geq$6}	&	35678	&	37031	&	579760	\\
 \hline
 	&	$\sim$5\%	&	$\sim$6\%	&	$\sim$89\%	\\
   \hline
 \end{tabular}
\caption{Events counts  after applying the selection cuts on the Geant output of the NH 500years data files in E$_\nu$=\{0.8,20\}GeV. }
\label{Table7}
\end{table}


 \section{ Contribution of the muonless events to $\nu$ Mass Hierarchy}
 In this section, we study effect of muonless events in the mass hierarchy determination. This is expected to be much smaller compared to that on the muon events. Nevertheless, we pursued this study with the hope of improving the hierarchy sensitivity of ICAL.
  \subsection{Physics Motivation and Application:}
  The matter effect modifies neutrino oscillation probabilities. For long pathlengths $(L \geq 5000\ {\rm km})$  and moderately large energies $(5\ {\rm GeV} \leq E_\nu \leq 10\ {\rm GeV})$, matter effects lead to large changes in both $P(\nu_\mu \to \nu_\mu)$ ($P_{\mu \mu}$) and $P(\nu_e \to \nu_\mu)$
($P_{e\mu}$). These changes can lead to an observable change in the muon event rate.
By measuring this change, it is possible to determine the neutrino mass hierarchy. The oscillation probabilities involving
$\nu_e$, $P(\nu_e \to \nu_e)$ ($P_{ee}$) and $P(\nu_\mu \to \nu_e)$ ($P_{\mu e}$), also undergo large changes due to matter
effects. The spectrum of the electron events is given by 
$$
\frac{dN_e}{d E_\nu} = \left[ \frac{d \Phi_e}{d E_\nu} P_{ee} + \frac{d \Phi_\mu}{d E_\nu} P_{\mu e} \right] \sigma_\nu.
$$
Since muon neutrino flux $d\Phi_\mu/dE_\nu$ is twice the electron neutrino flux $d \Phi_e/d E_\nu$ and the change in 
$P_{\mu e}$ is half the change in $P_{ee}$, the effect of these large changes mostly cancel each other out in the 
electron event sample. This fact makes finding matter effects in muonless events even more challenging.

 \subsection{The Generated Events Sample: }

The data files from Nuance, in the energy range $E_\nu$=\{0.1,100\}GeV are fed into the Geant4 INO ICAL code to get the events sample for the following studies. 
The neutrino oscillations have been applied  using the normal and the inverted mass hierarchy parameters, which are denoted as NH and IH respectively. The oscillation parameters used are as follows: $\Delta{{m}_{21}}^2 = 7.5\times 10^{-5}$ eV$^2$, $|\Delta{{m}_{eff}}^2| = 2.47\times 10^{-3}$ eV$^2$, $\sin^2\theta_{12}$ = 0.31, $\sin^2{2\theta_{13}}$ = 0.09, $\sin^2{\theta_{23}}$ = 0.5 and $\delta_{CP}$ = 0.
 
 \subsection{The Average MH $\chi^2$:}
\paragraph{}
Since we are using Nuance Monte-Carlo to calculate the number of events, we need to take
into account the MC fluctuations. Thus, if we simulate the NH events twice, with two
different seeds, the $\chi^2$ between these two event samples will be non-zero. In fact,
such $\chi^2_{\rm true} = \chi^2(NH1-NH2)$ will be approximately twice the number of bins. In addition,
we calculate $\chi^2_{\rm false} = \chi^2(IH-NH)$. If the NH is the true hierarchy,
then we expect $\chi^2_{\rm false}$ to be aprreciably greater than $\chi^2_{\rm true}$.

To minimize the overall effect of MC fluctuations, we do our calculations for very large
statistics and scale them down to 10 years. Here we consider data for 500 years. We have simulated
the data for NH with three different seeds and similarly for IH. Thus, we have six values
of $\chi^2({\rm true})$ and nine values of $\chi^2({\rm false})$. We take the average of each
and define the average $\chi^2$ for hierarchy as 
$$
 <\chi^2> = <\chi^2_{false}> - <\chi^2_{true}>.
$$
The numbers of the $\nu_e$CC and the $\nu_\mu$CC events generated for each of the seeds, are listed in Table~\ref{Table8}. 

	\begin{table}[H]		
 \centering
\begin{tabular}{|p{2.2cm}|p{2.2cm}|p{2.2cm}|p{2.2cm}|p{2.2cm}|}
 \hline
  \textbf{Sample ID}&{\textbf{NH $\nu_e$CC}}&{\textbf{NH $\nu_\mu$CC}}&{\textbf{IH $\nu_e$CC}}&{\textbf{IH $\nu_\mu$CC}}\\
 \hline
  \textbf{seed 1}	&676014		&1103263	&	671309	&	1103667\\
 \hline
  \textbf{seed 2}	&674971		&1103879	&	670827	&	1105891\\
 \hline
  \textbf{seed 3}	&675963		&1102817	&	669664	&	1104746\\
 \hline
 \end{tabular}
\caption{Events counts of $\nu_e$CC and $\nu_\mu$CC out of the 500 years Nuance data files, before interacting with the ICAL detector. Here, only the energy range {$E_\nu$=\{0.8,20\} GeV} is mentioned, which makes the major contribution in the value of $\chi^2$.}
\label{Table8}
\end{table}

 \subsection{Calculation of Average $\chi^2$ assuming Normal Hierarchy (NH):}
  The events are simulated in the energy range {$E_\nu$=\{0.1,100\} GeV}, for both NH and IH, each with three different seeds. We use the earlier mentioned criteria to select $\nu_e$CC rich samples in each case. To compare the distributions of these events in different cases, we sort them into a number of bins. We consider four different binning schemes. They are:
 
 \begin{itemize}
  \item 1-bin scheme: The events are all contained in one single bin. Each of these events must have a minimum of 11 hits and are confined in 4 or less number of layers. 
  \item 3-bin scheme: The selected sample  is divided into 3 bins, based on the number of hits. The events in the first bin should have a minimum of 11 hits but $\leq$ 20 hits. The second bin covers the range of 21 to 40 hits, while events with 40 to 100 hits are put in the third bin. Appropriate layers or hpl cuts are also included in each of them, so that the sample contains maximum number of $\nu_e$CC events.
  \item 10-bin scheme: The events are classified into 10 uniformly divided bins in the hits range 11 to 100, along with suitable  layers or hpl cuts.
  \item 15-bin scheme: The hits range 11 to 20 is divided into 10 bins. The events giving hits from 21 to 40 are grouped under 4 uniform bins. The fifteenth bin comprises of the events giving more than 40 hits. The layers or hpl cuts are imposed to ensure the majority of the $\nu_e$CC events in the sample.
 \end{itemize}

 We expect the  $<\chi^2>$ to increase with the increasing number of bins, until a saturation value is reached.

 The values for $\chi^2_{true}$ are calculated and tabulated in Table~\ref{Table9}, where as Table~\ref{Table10} contains the values of $\chi^2_{false}$.

	\begin{table}[H]		
 \centering
\begin{tabular}{|p{2cm}|c|c|c|c|c|}
 \hline
  \textbf{\small Sample pairs}&{\textbf{$\chi^2_{t}$} \small (1)}&{\textbf{$\chi^2_{t}$} \small (3 )}&{\textbf{$\chi^2_{t}$} \small (10 )}&{\textbf{$\chi^2_{t}$}\small  (15 )} \\
 \hline
  \textbf{\small NH2-NH1}	 	&	2&	6&	10&	30 \\
 \hline
  \textbf{\small NH3-NH1}	 	&	1&	3&	9&	27 \\
 \hline
  \textbf{\small NH3-NH2}	 	&	1&	9&	18&	17 \\
 \hline
 \hline
  \textbf{\small IH2-IH1}	 	&	3&	9&	14&	44 \\
 \hline
  \textbf{\small IH3-IH2}	 	&	0&	14&	23&	22 \\
 \hline
  \textbf{\small IH3-IH1}	 	&	2&	7&	19&	47 \\
 \hline
 \hline
\textbf{\small Average }	 	&	2&	8&	16&	31 \\	
 \hline
 \end{tabular}
\caption{Values of $\chi^2_{true}$ or $\chi^2_{t}$ from the three possible combinations of the three sets of dataset assuming Normal Mass Hierarchy for the binning schemes: (a) 1 bin-scheme; (b) 3 bin-scheme; (c) 10 bin-scheme; (d) 15 bin-scheme.}
\label{Table9}
\end{table}

\begin{table}[H]		
 \centering
\begin{tabular}{|p{2cm}|c|c|c|c|}
 \hline
  \textbf{\small Sample pairs}&{\textbf{$\chi^2_{f}$} \small ( 1 )}&{\textbf{$\chi^2_{f}$} \small ( 3 )}&{\textbf{$\chi^2_{f}$} \small ( 10 )}&{\textbf{$\chi^2_{f}$} \small ( 15 )} \\
 \hline
  \textbf{\small NH1-IH1}	 	&	44	&	55&	72	&	82	\\
 \hline
  \textbf{\small NH1-IH2}	 	&	26	&	29&	46	&	62	\\
 \hline
  \textbf{\small NH1-IH3}	 	&	39	&	39&	50	&	70	\\
 \hline
  \textbf{\small NH2-IH1}	 	&	27	&	29&	43	&	50	\\
 \hline
  \textbf{\small NH2-IH2}	 	&	13	&	14&	31	&	63	\\
 \hline
  \textbf{\small NH2-IH3}	 	&	23	&	28&	39	&	54	\\
 \hline
  \textbf{\small NH3-IH1}	 	&	35	&	53&	80	&	74	\\
 \hline
  \textbf{\small NH3-IH2}	 	&	19	&	23&	45	&	67	\\
 \hline
  \textbf{\small NH3-IH3}	 	&	30	&	33&	50	&	60	\\
 \hline
 \hline
  \textbf{\small Average (all) }	 &	28	&	34&	51	&	65	\\
 \hline
 \end{tabular}
\caption{Values of $\chi^2_{false}$ or $\chi^2_{f}$ from the nine possible combinations of the three sets of dataset assuming/expecting Normal Mass Hierarchy. (a) 1 bin-scheme; (b) 3 bin-scheme; (c) 10 bin-scheme; (d) 15 bin-scheme.}
\label{Table10}
\end{table}

 In Table~\ref{Table11}, we have listed $<\chi^2> = <\chi^2_{false}> - <\chi^2_{true}>$ and the standard deviation in $<\chi^2>$.
 This standard deviation is simply the sum of the standard deviations from the mean values of $\chi^2_{true}$ and $\chi^2_{false}$.


	\begin{table}[H]		
 \centering
\begin{tabular}{|c|c|c|c|}
 \hline
  \textbf{hits  binning}&{\textbf{$<\chi^2>$} 500yrs}&{\textbf{$\sigma(<\chi^2>)$ or $\chi^2_E$}500 yrs}&\textbf{$<\chi^2>$} in 10 yrs \\
 \hline
  \textbf{1 }	 	&	26	&	10&	0.5 \\
 \hline
  \textbf{3 }	 	&	26	&	16&	0.5 \\
 \hline
  \textbf{10 }	 	&	35	&	20&	0.7 \\
 \hline
  \textbf{15 }	 	&	34	&	20&	0.7 \\
 \hline
  

 \end{tabular}
\caption{Values of average $\chi^2$ with standard deviation as the error in varying number of bins. The values  are also scaled down to 10 years. }
\label{Table11}
\end{table}

A cos$\theta$ binning has been attempted, but it appears to be of no additional  help in the present study. The binning in terms of number of hits seems to suffice.

In the above calculations, we assumed that NH is true. The results for the case IH is true are 
very similar.

 \subsection{The Frequentist Approach with 10 years events samples}
 
 The statistical fluctuations in the data are {intrinsic} to  it. The Monte Carlo simulations mimic these fluctuations but every different simulation carries with it a different set of fluctuations. To properly estimate the effect of statistical fluctuations on $<\chi^2>$, one should do a number of NH and IH simulations. From these, one can obtain the average value and the standard deviation of $\chi^2_{true}$ and $\chi^2_{false}$. The difference in the average values of $\chi^2_{true}$ and $\chi^2_{false}$ gives the expected mean value of the $<\chi^2>$ and the sum of their standard deviations gives us the probable range of $<\chi^2>$ \cite{Blennow:2013oma}.

We have divided the 500 years of data into fifty 10 years data samples, for both NH and IH. Using these, we computed 1225 values of NH-NH $\chi^2_{true}$ and 2500 values of NH-IH $\chi^2_{false}$. Their distributions are given in figure 6. The $<\chi^2_{false}>$ is higher than the $<\chi^2_{true}>$ by almost 3 (i.e. $<\chi^2>$ $\sim$3). 

 	\begin{figure}[H]	
 \centering
 \setlength\fboxsep{0pt}
 {\includegraphics[width=0.95\textwidth ]{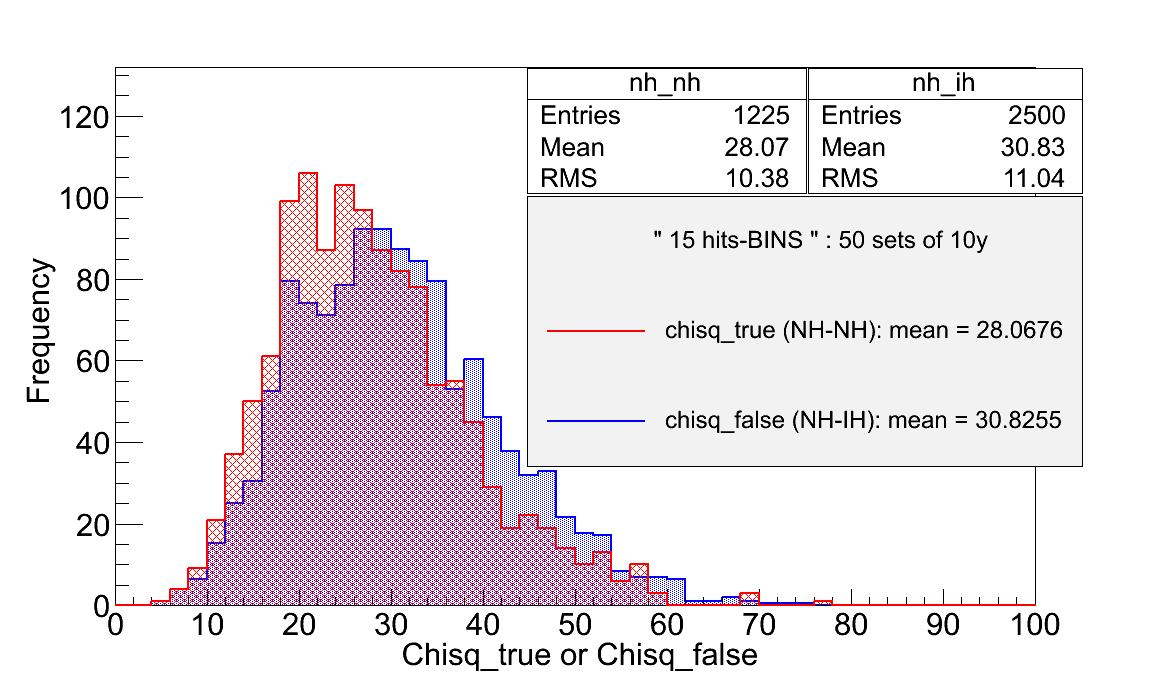}}
 \caption{Distribution of $\chi^2_{\rm false}$ \& $\chi^2_{\rm true}$ calculated with 10years of NH data and IH data, assuming NH ordering to be true.   }
\end{figure}

%

The plot in figure 6 show the $\chi^2_{\rm false}$ and $\chi^2_{\rm true}$ distributions in the 15-bin scheme. However, both the $\chi^2_{\rm false}$ and the $\chi^2_{\rm true}$ have very broad distributions. Hence, one is not able to make any conclusive statement on the $<\chi^2>$.

%

\section{Summary and Conclusion}

 The hits and layers cuts have been confirmed to be the most important criteria for selecting a $\nu_e$CC rich events sample. They alone fetch us an event sample containing around $\sim$ 50\% $\nu_e$CC events.  The selection criteria to be finally chosen depend on the requirements of the physics study. One might insist on the maximum possible purity of 
 the $\nu_e$CC events, even compromising the vertical events fraction or small sample size.

 
 The effects of the various other additional selection criteria are tabulated in Table~\ref{Table12}.
 
 \begin{table}[H]		
 \centering
\begin{tabular}{|p{4.5cm}|p{2.5cm}|p{2.4cm}|p{2.5cm}|}

 \hline
  \textbf{Selection Criteria}&\textbf{\small Best Ratio of \#$\nu_e$CC to total \#events}&\textbf{Max. sample size with the ratio (k=10$^3$)}&\textbf{Remarks}\\
 \hline
   \textbf{Maximum Hits diff.}			&  53\%	&	156k	&{\footnotesize Large sample size}\\
 \hline
  \textbf{Comparison: hits in layers}		&  60\%	&	43k	&\\
 \hline
  \textbf{Overall Pattern: hits in layers}	&  62\%(58\%)	&	26k(88k)	&\\
 \hline
   \textbf{Single layer hits}			&  68\%	&	6.5k	&{\footnotesize Very small sample size; Single layer more prone to noise}\\
 \hline
 
\end{tabular}
\caption{Effects of different selection criteria on the event sample after putting cuts on \#hits and layers. Purity Ratio of $\nu_e$CC in the total sample, and the corresponding sample size for 500 years NH data are shown.}
\label{Table12}
\end{table}

The results of optimizing the selection criteria may also be summed up in the following manner:
 \begin{itemize}
  \item Maximum obtainable {\bf $\nu_e$CC purity: $\sim$ 60\%} with counts $\sim$ 100 events per year or {\bf $\sim$ 62\%} with counts $\sim$ 50 events per year.
  \item Maximum obtainable {\bf $\nu$NC purity: $\sim$ 47\%} with counts $\sim$ 1800 events per year, provided there is no noise.
  \item The selection criteria described in this report retain a majority of horizontal or near horizontal events. Any efforts to increase the percentage of vertical or near vertical events requires a compromise on the part of purity of the $\nu_e$CC events.
 \end{itemize}

 A Nuance based analysis of the selected events has been done to understand the composition of the selected events. Three types of interactions are studied. The $\nu_e$CC events chosen by selection cuts are thrice the background in the QE interaction bin. The NC and the $\nu_\mu$CC sum up to just half the number of $\nu_e$CC events in the RS bin. The DIS events count of the $\nu_e$CC  almost equals the sum of other two.  This implies, that the selection cuts will distinguish the QE and RS $\nu_e$CC events over the rest, but not the DIS events. But the number of QE events are very less in comparison to the other two types. So, the overall signal to background ratio is not so good as can be expected.

  In the selection of $\nu_\mu$CC events sample, $\nu_e$CC + NC events form  a 20\% background, if the events have a minimum of 5 layers hit. If this minimum is raised to 6 layers, the background contribution comes down to 10\%.
  We also conclude that the contribution of the muonless events in determining the neutrino mass hierarchy is not zero. But the statistical fluctuations in the data are too large for this contribution to have a significant effect.
 
\paragraph{Acknowledgement}
\paragraph{}
 We express our deep gratitude to all our co-members of the INO Collaboration. We particularly thank Prof. Gobinda Majumder and Prof. Indumathi for their invaluable suggestions. We also specially thank Lakshmi S. Mohan for the useful discussions involving her work on separating $\nu_\mu$CC events from the NC events. We extend our thanks to Tarak Thakore for numerous discussions on Nuance and calculations of mass hierarchy $\chi^2$. We are grateful for the worthy suggestions from Prof. Amol Dighe regarding the mass hiararchy contribution of the muonless events. We also thank the Department of Atomic Energy (DAE) and the Department of Science and Technology, Government of India, for financial support. 

 \bibliographystyle{unsrt}
 \bibliography{SML_bibtex_inspire}
 
\end{document}